\documentclass[pra,floatfix,amsmath,twocolumn,showpacs,superscriptaddress]{revtex4}
\usepackage[ansinew]{inputenc}
\usepackage{times}
\usepackage{verbatim}
\usepackage{epsfig}
\usepackage{pst-all}
\usepackage{color}
\usepackage{dcolumn}
\usepackage{amsmath}
\usepackage{amsthm}
\usepackage{bm}
\usepackage{layout}
\usepackage{float}
\usepackage{txfonts}
\usepackage{amsfonts}
\usepackage{amssymb}%
\usepackage{subfigure}
\usepackage[ansinew]{inputenc}
\usepackage{amsmath}
\usepackage{amsthm}
\usepackage{float}
\usepackage{amsfonts}
\usepackage{amssymb}

\begin{document}
\title{Multipartite distribution property of one way discord beyond measurement}

\author{Si-Yuan Liu }
\affiliation{Institute of Modern Physics, Northwest University, Xian
710069, P. R. China }
\affiliation{Beijing National Laboratory for Condensed Matter Physics,
Institute of Physics, Chinese Academy of Sciences, Beijing 100190, P. R. China}

\author{Yu-Ran Zhang}
\affiliation{Beijing National Laboratory for Condensed Matter Physics,
Institute of Physics, Chinese Academy of Sciences, Beijing 100190, P. R. China}

\author{Wen-Li Yang }
\email{wlyang@nwu.edu.cn }
\affiliation{Institute of Modern Physics, Northwest University, Xian
710069, P. R. China }

\author{Heng Fan}
\email{hfan@iphy.ac.cn}
\affiliation{Beijing National Laboratory for Condensed Matter Physics,
Institute of Physics, Chinese Academy of Sciences, Beijing 100190, P. R. China}
\affiliation{Collaborative Innovation Center of Quantum Matter, Beijing, P. R. China}

\date{\today}

\begin{abstract}
We investigate the distribution property of one way discord in multipartite system
by introducing the concept of polygamy deficit for one way discord.
The difference between one way discord and quantum discord is analogue to the
difference between entanglement of assistance and entanglement of formation.
For tripartite pure states, two kinds of polygamy deficits are presented with
the equivalent expressions and physical interpretations regardless of measurement.
For four-partite pure states, we provide a condition which makes one way discord polygamy
being satisfied. Those results can be applicable to multipartite quantum systems
and are complementary to our understanding of the shareability of quantum correlations. 
\end{abstract}


\maketitle

\section{introduction}
Quantum correlations, such as entanglement and quantum discord, are
considered as valuable resources for quantum information task \cite{key-1,key-2,key-3,key-4,key-5,key-6,key-7}.
On the other hand, in general, entanglement and discord are quite
different from each other. Since the entanglement seems not to capture
all the quantum features of quantum correlations, other measures of
quantum correlations are proposed. Quantum discord is a widely accepted
one among them \cite{key-8,key-9,key-10,key-11,key-12,key-13,key-14,key-15}.
The quantum discord plays an important role in the research of quantum
correlations due to its potential applications in a number of quantum
processes, such as quantum critical phenomena \cite{key-16,key-17,key-18,key-19},
quantum evolution under decoherence \cite{key-20,key-21,key-22} and
the DCQ1 protocol \cite{key-23}.

Since quantum discord quantifies the quantum correlation in a bipartite
state and might also be a resource in quantum-information processing,
it is interesting to study its distribution property in the multipartite
system. The monogamy property which characterizes the restriction
for sharing a resource or a quantity is helpful to provide significant
information for this issue and deserves systematic investigation.
In general, the limits on the shareability of quantum correlations
are described by monogamy inequalities \cite{key-24,key-25,key-26,key-27,key-28,key-29}.
Recently, the monogamy relation for quantum discord was studied in \cite{key-30,key-31,key-32,key-33,key-34,key-35,key-36}. It is found
that the monogamy of quantum discord is not always hold for any tripartite
pure state \cite{key-33}. That is to say, the polygamy relation for
quantum discord can hold for some states.

Recently, a quantum correlation similar as quantum discord called
one-way unlocalizable quantum discord was presented \cite{key-37}.
The one way discord has an operational interpretation, for any tripartite
pure state, the polygamy relation always holds. It is interesting
to study the difference and connection between one way discord and
quantum discord. The distribution property of one way discord in multipartite
system is also worth considering. In this paper, we present the concept
of polygamy deficit of one way discord. For any tripartite pure state,
using the equivalent expression of polygamy deficit, we can control
the polygamy degree of one way discord. For 4-partite pure states,
we provide an condition for the case that one way discord is polygamy.
We believe that our results provide a useful method in understanding
the distribution property of one way discord.
Our results get rid of the optimal measurement problem,
which is difficult or even impossible to overcome in most researches on
one way discord and quantum discord, and give important relations to simplify the calculation,
Therefore, we believe that our results may have great applications in quantum information
processing  and can be applied to physical models of many-body quantum
systems.

The paper is organized as follows. In Sec. II, we give a brief review
of the definition of one way discord and corresponding correlations.
In Sec. III, we study the differences and connections between one
way discord, quantum discord and corresponding quantum correlations.
In Sec. IV, we define the polygamy deficit of one way discord. For
any tripartite pure state, the polygamy degree of one way discord
is considered. For 4-partite pure states, we provide an condition that makes
one way discord polygamy. 
In Sec. V, we summarize our results.

\section{The definition of one way discord and corresponding correlations\label{II}}
In order to study the distribution property of one way discord, we
give a brief review of one way discord and corresponding correlations.

For a bipartite state $\rho_{AB}$, the one-way unlocalizable quantum
discord is defined as the difference between the mutual information
and the one way unlocalizable entanglement \cite{key-37}, namely,
\begin{eqnarray}
\delta_{u}^{\leftarrow}\left(\rho_{AB}\right)=I\left(\rho_{AB}\right)-E_{u}^{\leftarrow}\left(\rho_{AB}\right),
\end{eqnarray}
where $I\left(\rho_{AB}\right)=S\left(\rho_{A}\right)+S\left(\rho_{B}\right)-S\left(\rho_{AB}\right)$
is mutual information \cite{key-38}. The one way unlocalizable entanglement (UE)
is defined as:
\begin{eqnarray}
E_{u}^{\leftarrow}\left(\rho_{AB}\right)=\underset{\left\{ M_{k}^{B}\right\} }{\min}\left[S\left(\rho_{A}\right)-\underset{k}{\sum}p_{k}S(\rho_{k}^{A})\right],
\end{eqnarray}
where the minimum is taken over all possible rank-1 measurements $\{ M_{k}^{B}\} $
applied on subsystem $B$, $p_{k}=\textrm{Tr}[(I^{A}\otimes M_{k}^{B})\rho^{AB}]$
is the probability of the outcome $k$, and $\rho_{k}^{A}=\textrm{Tr}_{B}[(I^{A}\otimes M_{k}^{B})\rho_{AB}]/p_{k}$
is the state of system $A$ when the outcome is $k$ \cite{key-39}.
The definition of quantum discord is similar as one way discord,
\begin{eqnarray}
D^{\leftarrow}\left(\rho_{AB}\right)=I\left(\rho_{AB}\right)-J^{\leftarrow}\left(\rho_{AB}\right),
\end{eqnarray}
where $J^{\leftarrow}\left(\rho_{AB}\right)$ is the classical correlation
which defined as
\begin{eqnarray}
J^{\leftarrow}\left(\rho_{AB}\right)=\underset{\left\{ M_{k}^{B}\right\} }{\max}\left[S\left(\rho_{A}\right)-\underset{k}{\sum}p_{k}S(\rho_{k}^{A})\right].
\end{eqnarray}
For any tripartite pure state, we have the Koashi-Winter relation \cite{key-27}
\begin{eqnarray}
J^{\leftarrow}\left(\rho_{AB}\right)+E_{f}\left(\rho_{AC}\right)=S\left(\rho_{A}\right),\label{5}
\end{eqnarray}
where $E_{f}\left(\rho_{AC}\right)$ is the entanglement of formation (EOF) of $\rho_{AC}$.
Similarly, we have the Buscemi-Gour-Kim equality \cite{key-39}:
\begin{eqnarray}
E_{u}^{\leftarrow}\left(\rho_{AB}\right)=S\left(\rho_{A}\right)-E_{a}\left(\rho_{AC}\right),\label{6}
\end{eqnarray}
where $E_{a}\left(\rho_{AC}\right)$ is the entanglement of assistance (EOA)
of $\rho_{AC}$, which is defined by the maximum average entanglement
of $\rho_{AC}$ \cite{key-40,key-41},
\begin{eqnarray}
E_{a}\left(\rho_{AC}\right)=\underset{\left\{ p_{x},\ |\phi_{x}\rangle^{AC}\right\} }{\max}\underset{x}{\sum}p_{x}S(\rho_{x}^{A}),
\end{eqnarray}
where the maximum is taken over all possible pure-state decompositions
of $\rho_{AC}$, satisfying $\rho_{AC}=\sum_{x}p_{x}|\phi_{x}\rangle^{AC}\langle\phi_{x}|$
and $\rho_{x}^{A}=\textrm{Tr}_{C}(|\phi_{x}\rangle^{AC}\langle\phi_{x}|)$.
Here, $S\left(\rho\right)=-\textrm{Tr}(\rho\log_{2}\rho)$ is the von Neumann
entropy \cite{key-38}.

\section{The difference and connection between one way discord and quantum discord\label{III}}
The one way discord and quantum discord are two similar quantum correlations. The differences and connections between them are interesting questions. In this section, we study this issue carefully. First of all, let us consider the difference between one way discord and quantum discord. According to the definitions, we have
$
D^{\leftarrow}\left(\rho_{AB}\right)+J^{\leftarrow}\left(\rho_{AB}\right)=\delta_{u}^{\leftarrow}\left(\rho_{AB}\right)+E_{u}^{\leftarrow}\left(\rho_{AB}\right)=I\left(\rho_{AB}\right),
$
Using Eq.~(\ref{5}) and (\ref{6}), we have
\begin{eqnarray}
\delta_{u}^{\leftarrow}\left(\rho_{AB}\right)-D^{\leftarrow}\left(\rho_{AB}\right)=E_{a}\left(\rho_{AC}\right)-E_{f}\left(\rho_{AC}\right)\geq0.
\end{eqnarray}
This formula tells us an interesting fact that the difference between
one way discord and quantum discord for $\rho_{AB}$ is equivalent
to the difference between EOA and EOF for $\rho_{AC}$ .

We know that
the one way discord is greater than or equal to the quantum discord
in general, but how to measure the difference between them? For an
arbitrary tripartite pure state, this equation tells us that the difference
can be measured by the difference between EOA and EOF for another
two parties. Since the EOF and EOA do not contains measurements, it
is much easier to calculate the difference between them, which
provides an simple method for measureing the difference between one
way discord and quantum discord.  If $\rho_{ABC}$ changes from a
pure state to another pure state, the change of both sides of this
equation are equivalent.  In other
words, we can control the difference between this two kinds of quantum
discord by adjusting the corresponding entanglement measure. In particular,
when EOA and EOF are equal, the two kinds of quantum discord are equivalent.

\begin{figure}[t]
\centering
\subfigure[]{\includegraphics[width=0.22\textwidth]{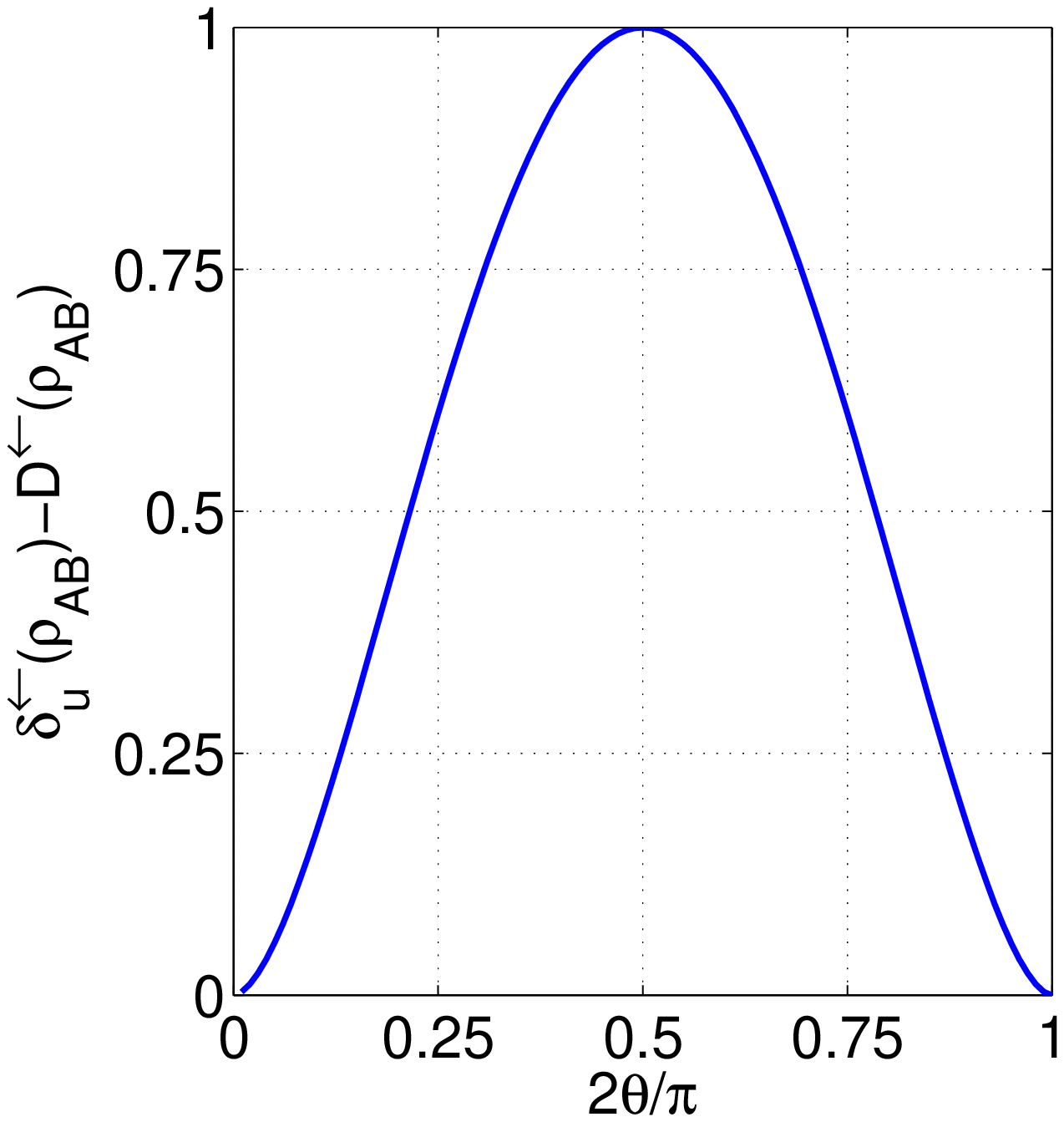}}
\subfigure[]{\includegraphics[width=0.22\textwidth]{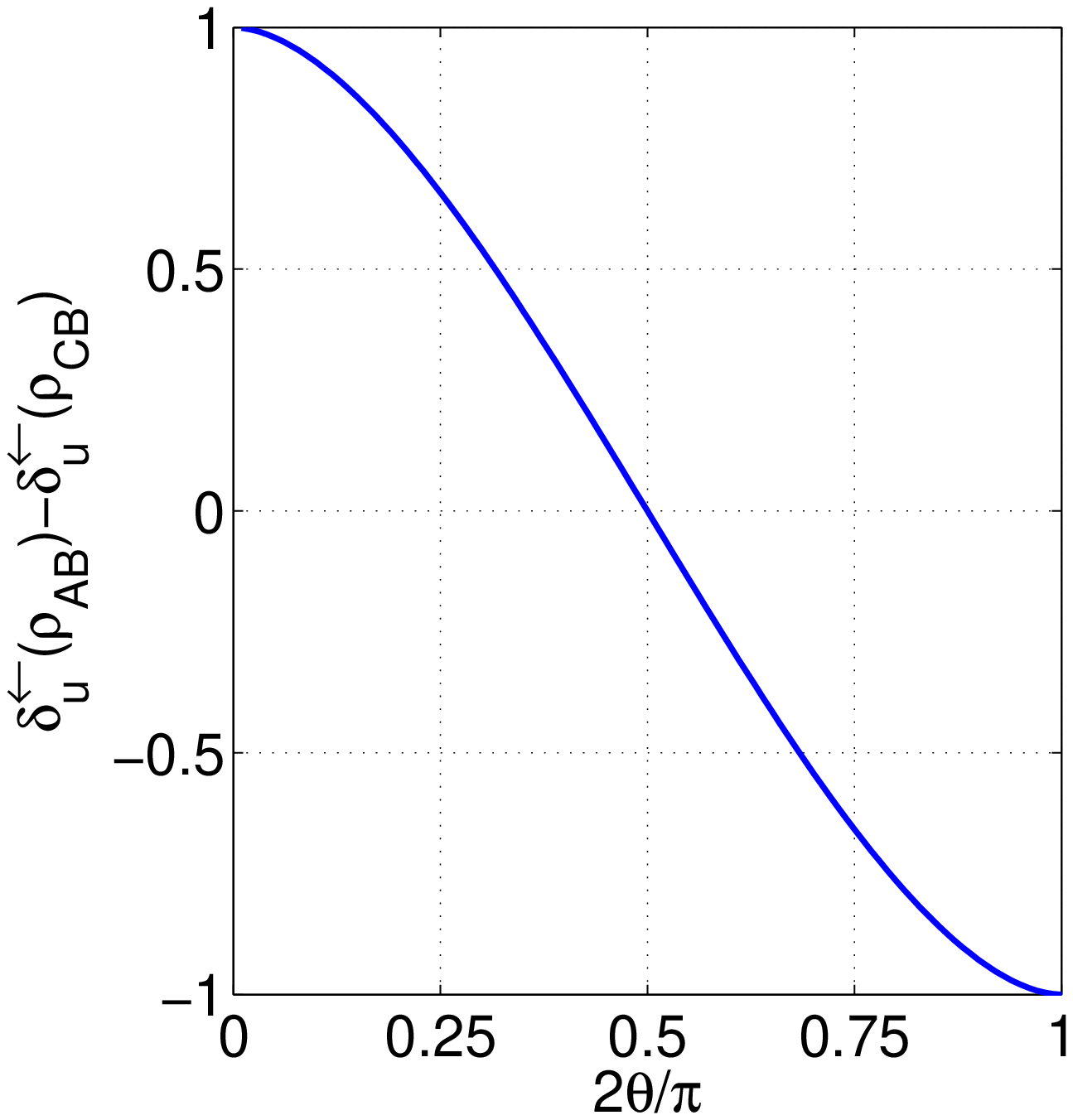}}
\caption{(color online). (a) $\delta_{u}^{\leftarrow}\left(\rho_{AB}\right)-D^{\leftarrow}\left(\rho_{AB}\right)$
v.s. ${2\theta}/{\pi}$ for the GHZ state $|\psi\rangle=\cos\theta|000\rangle+\sin\theta|111\rangle$.
(b) The difference between one way discord or quantum discord
of $\rho_{AB}$ and $\rho_{CB}$ is plotted as a function of ${2\theta}/{\pi}$ for a state
$|\psi\rangle=\cos\theta|\Phi^{+}\rangle_{AB}|0\rangle_{C}+\sin\theta|0\rangle_{A}|\Psi^{+}\rangle_{BC}$.}
\label{f1}
\end{figure}

Now we give a simple example. In Fig.~1(a), for the GHZ state $|\psi\rangle=\cos\theta|000\rangle+\sin\theta|111\rangle$ ($\theta\in\left[0,{\pi}/{2}\right]$), the $\delta_{u}^{\leftarrow}\left(\rho_{AB}\right)-D^{\leftarrow}\left(\rho_{AB}\right)$
is plotted as a function of ${2\theta}/{\pi}$. This figure
shows that the difference between one way discord and quantum discord
for $\rho_{AB}$ first increases then decreases with increasing $\theta$.
In particular, when $\theta=$ or ${\pi}/{2}$, the state is separable and we have $\delta_{u}^{\leftarrow}\left(\rho_{AB}\right)=D^{\leftarrow}\left(\rho_{AB}\right)$.
When $\theta$= ${\pi}/{4}$, the state is the maximally entangled
state, the difference between one way discord and quantum discord
for $\rho_{AB}$ reaches the maximum value.

Consider the distribution property of one way discord, we provide
an interesting relationship as follows:
\begin{eqnarray}
\delta_{u}^{\leftarrow}\left(\rho_{AB}\right)-\delta_{u}^{\leftarrow}\left(\rho_{CB}\right)=D^{\leftarrow}\left(\rho_{AB}\right)-D^{\leftarrow}\left(\rho_{CB}\right)
=S\left(B|C\right).
\label{11}
\end{eqnarray}
Here we give a simple proof. Using the definition and Buscemi-Gour-Kim equality, we have
$\delta_{u}^{\leftarrow}\left(\rho_{AB}\right)=I\left(\rho_{AB}\right)-E_{u}^{\leftarrow}\left(\rho_{AB}\right)$, with
$E_{u}^{\leftarrow}\left(\rho_{AB}\right)=S\left(\rho_{A}\right)-E_{a}\left(\rho_{AC}\right)$.
Thus, we have
$\delta_{u}^{\leftarrow}\left(\rho_{AB}\right)=E_{a}\left(\rho_{AC}\right)-S\left(A|B\right)$ and
$\delta_{u}^{\leftarrow}\left(\rho_{CB}\right)=E_{a}\left(\rho_{CA}\right)-S\left(C|B\right)$, similarly.
Combined these two equations, we have
\begin{eqnarray}
\delta_{u}^{\leftarrow}\left(\rho_{AB}\right)-\delta_{u}^{\leftarrow}\left(\rho_{CB}\right)=S\left(\rho_{A}\right)-S\left(\rho_{C}\right)=S\left(B|C\right).
\end{eqnarray}
For quantum discord, using the Koashi-Winter equality, we have
$D^{\leftarrow}\left(\rho_{AB}\right)=E_{f}\left(\rho_{AC}\right)-S\left(A|B\right)$,
and $
D^{\leftarrow}\left(\rho_{CB}\right)=E_{f}\left(\rho_{CA}\right)-S\left(C|B\right)$.
The difference between $D^{\leftarrow}\left(\rho_{AB}\right)$ and
$D^{\leftarrow}\left(\rho_{CB}\right)$ is equivalent to $S\left(\rho_{A}\right)-S\left(\rho_{C}\right)$,
which completes the proof.

This equation gives us another interesting fact that the difference between
one way discord of $\rho_{AB}$ and $\rho_{CB}$ is equivalent to
the difference between quantum discord of the same states. Both of
them are equal to the conditional entropy $S\left(B|C\right)$
or $-S\left(B|A\right)$. That is to say, the difference is independent
of measure and the quantum correlations we used. For any tripartite
pure states, if $S\left(\rho_{A}\right)$ is greater than or equal
to $S\left(\rho_{C}\right)$, we have the conditional entropy $S\left(B|C\right)\geq0$, which means
the one way discord or quantum discord of $\rho_{AB}$
is always greater than or equal to $\rho_{CB}$, and vice versa. This
formula also gives a new physical meaning for the conditional entropy:
it reflects the distribution property of one way discord or quantum
discord for relevant states. It tells us that if we want to control
the distribution of one way discord or quantum discord between $\rho_{AB}$
and $\rho_{CB}$, we only need to adjust the corresponding conditional
entropy.
If $\rho_{ABC}$ changes from a pure state to another pure
state, the change of both sides of this equation are equivalent.

For example, we consider a state,
$
|\psi\rangle=\cos\theta|\Phi^{+}\rangle_{AB}|0\rangle_{C}+\sin\theta|0\rangle_{A}|\Psi^{+}\rangle_{BC}
$
with $|\Psi^{+}\rangle$ and $|\Phi^{+}\rangle$ the Bell states.
In Fig.~1(b), the difference between one way discord or quantum discord
of $\rho_{AB}$ and $\rho_{CB}$ is plotted as a function of ${2\theta}/{\pi}$ ($\theta=[0,{\pi}/{2}]$). This figure shows that
the difference between one way discord or quantum discord of $\rho_{AB}$
and $\rho_{CB}$ decreases from $1$ to $-1$ with increasing $\theta$.
In particular, when $\theta$= ${\pi}/{4}$, the conditional entropy
$S\left(B|C\right)$ is equal to zero, the one way discord or
quantum discord of $\rho_{AB}$ and $\rho_{CB}$ are equal.

\section{The polygamy deficit of one way discord\label{IV}}
In order to study the distribution property of one way discord carefully, similar as the quantum
discord, we give two kinds of polygamy deficits of one way discord \cite{key-35},
\begin{eqnarray}
\triangle_{\delta_{u\left(A\right)}}^{\leftarrow}=\delta_{u}^{\leftarrow}\left(\rho_{A\left(BC\right)}\right)-\delta_{u}^{\leftarrow}\left(\rho_{AB}\right)-\delta_{u}^{\leftarrow}\left(\rho_{AC}\right).
\\
\triangle_{\delta_{u\left(A\right)}}^{\rightarrow}=\delta_{u}^{\rightarrow}\left(\rho_{A\left(BC\right)}\right)-\delta_{u}^{\rightarrow}\left(\rho_{AB}\right)-\delta_{u}^{\rightarrow}\left(\rho_{AC}\right).
\end{eqnarray}
The $\triangle_{\delta_{u\left(A\right)}}^{\leftarrow}$  involves the local measurements on
$B$, $C$ and a coherent measurement on $BC$, while the $\triangle_{\delta_{u\left(A\right)}}^{\rightarrow}$ only involves local measurements on $A$.
The first kind of polygamy deficit can be rewritten as
\begin{eqnarray}
\triangle_{\delta_{u\left(A\right)}}^{\leftarrow}=E_{u}^{\leftarrow}\left(\rho_{AB}\right)-E_{a}\left(\rho_{AB}\right)=E_{u}^{\leftarrow}\left(\rho_{AC}\right)-E_{a}\left(\rho_{AC}\right).
\label{13}
\end{eqnarray}
Here we give a simple proof. For tripartite pure states, we have
$\triangle_{\delta_{u\left(A\right)}}^{\leftarrow}=S\left(\rho_{A}\right)-\delta_{u}^{\leftarrow}\left(\rho_{AB}\right)-\delta_{u}^{\leftarrow}\left(\rho_{AC}\right)$.
Using the formulas we have proved, the polygamy deficit can be re-expressed
as
\begin{eqnarray}
\triangle_{\delta_{u\left(A\right)}}^{\leftarrow}&=&S\left(\rho_{A}\right)-I\left(\rho_{AB}\right)+E_{u}^{\leftarrow}\left(\rho_{AB}\right)-E_{a}\left(\rho_{AB}\right)+S\left(A|C\right)
\nonumber\\
&=&E_{u}^{\leftarrow}\left(\rho_{AB}\right)-E_{a}\left(\rho_{AB}\right).
\end{eqnarray}
Using the Buscemi-Gour-Kim equality
$E_{u}^{\leftarrow}\left(\rho_{XY}\right)+E_{a}\left(\rho_{XZ}\right)=S\left(\rho_{X}\right)$  ($X,Y,Z\in\left\{ A,B,C\right\}$),
we have
$
E_{u}^{\leftarrow}\left(\rho_{AB}\right)-E_{a}\left(\rho_{AB}\right)=E_{u}^{\leftarrow}\left(\rho_{AC}\right)-E_{a}\left(\rho_{AC}\right)$.

According to Ref.~\cite{key-37}, the one way discord is polygamy
for tripartite pure states, but we do not know the degree of polygamy
for a particular state. Eq.~(\ref{13}) shows us
that the degree of polygamy for one way discord is determined by the
difference of UE and EOA. That is to say, we can control the degree
of polygamy by adjusting the difference of UE and EOA for corresponding
reduced state. The polygamy deficit can be reduced by decreasing the
difference of UE and EOA. It is worth noting that the right hand side
of this equation only contains one local measurement on $B$ or $C$, so
the experiment and calculations can be greatly simplified. The difference
of UE and EOA for $\rho_{AB}$ and $\rho_{AC}$ change in the same
step, we only need to consider one of them.

Now we consider the equivalent expression of the second polygamy deficit of one way discord. For tripartite pure states, we have
$\triangle_{\delta_{u\left(A\right)}}^{\rightarrow}=S\left(\rho_{A}\right)-\delta_{u}^{\rightarrow}\left(\rho_{AB}\right)-\delta_{u}^{\rightarrow}\left(\rho_{AC}\right)$,
where $\delta_{u}^{\rightarrow}\left(\rho_{AB}\right)=\delta_{u}^{\leftarrow}\left(\rho_{BA}\right),\delta_{u}^{\rightarrow}\left(\rho_{AC}\right)=\delta_{u}^{\leftarrow}\left(\rho_{CA}\right)$.
Since we have proved in the previous section that
$\delta_{u}^{\leftarrow}\left(\rho_{XY}\right)+S\left(X\mid Y\right)=E_{a}\left(\rho_{XZ}\right)$ with ($X,Y,Z\in\left\{ A,B,C\right\}$).
It can be rewritten as
\begin{eqnarray}
\triangle_{\delta_{u\left(A\right)}}^{\rightarrow}=S\left(\rho_{A}\right)+S\left(B|A\right)+S\left(C|A\right)-2E_{a}\left(\rho_{BC}\right).
\end{eqnarray}
For tripartite pure states, we have
\begin{eqnarray}
\triangle_{\delta_{u\left(A\right)}}^{\rightarrow}=I\left(\rho_{BC}\right)-2E_{a}\left(\rho_{BC}\right).
\end{eqnarray}
This equation tells us an interesting fact that the second polygamy
inequality also holds for one way discord, since $E_{a}\left(\rho_{BC}\right)\geq\frac{1}{2}I\left(\rho_{BC}\right)$
always holds for tripartite pure states. That is to say, both the
first polygamy inequality contains local and coherent measurements
and the second polygamy inequality only contains local measurement
hold for one way discord. The polygamy degree of the second polygamy
inequality is decided by the difference of $I\left(\rho_{BC}\right)$
and $2E_{a}\left(\rho_{BC}\right)$. In other words, we can control
the polygamy degree by adjusting the mutual Information and EOA for
$\rho_{BC}$. It is worth noting that the right hand side of this
equation does not include any measurement, so the experiment and calculations
can be greatly simplified.
We can also prove that
\begin{eqnarray}
\triangle_{\delta_{u\left(A\right)}}^{\rightarrow}=E_{u}^{\leftarrow}\left(\rho_{BA}\right)-\delta_{u}^{\leftarrow}\left(\rho_{BA}\right)=E_{u}^{\leftarrow}\left(\rho_{CA}\right)-\delta_{u}^{\leftarrow}\left(\rho_{CA}\right).
\end{eqnarray}
It  shows that the second polygamy
deficit is equivalent to the difference between UE and one way discord
for $\rho_{BA}$ or $\rho_{CA}$, which change in the same step. In
particular, when $E_{a}\left(\rho_{BC}\right)=\frac{1}{2}I\left(\rho_{BC}\right)$,
we have $\triangle_{\delta_{u\left(A\right)}}^{\rightarrow}=0$ and
$E_{u}^{\leftarrow}\left(\rho_{BA}\right)=\delta_{u}^{\leftarrow}\left(\rho_{BA}\right)$,
$E_{u}^{\leftarrow}\left(\rho_{CA}\right)=\delta_{u}^{\leftarrow}\left(\rho_{CA}\right)$.

Now we provide a simple example. In Fig.~2(a), for the GHZ state $|\psi\rangle=\cos\theta|000\rangle+\sin\theta|111\rangle$
($\theta\in\left[0,{\pi}/{2}\right]$), the $\triangle_{\delta_{u\left(A\right)}}^{\rightarrow}=I\left(\rho_{BC}\right)-2E_{a}\left(\rho_{BC}\right)$
is plotted as a function of $\frac{2\theta}{\pi}$. This figure
shows that the second polygamy deficit of one way discord first
decreases then increases with increasing $\theta$. In particular,
when $\theta=0$ or ${\pi}/{2}$, the state is separable state, we have
$\triangle_{\delta_{u\left(A\right)}}^{\rightarrow}=0$. When $\theta$=
${\pi}/{4}$, the state is maximally entangled state, the polygamy
degree of the second polygamy inequality reaches maximum, $\triangle_{\delta_{u\left(A\right)}}^{\rightarrow}=-1$.

Similarly, for 4-partite pure state, we define the two kinds of polygamy deficits as follows:
\begin{eqnarray}
\triangle_{\delta_{u\left(A\right)}}^{^{\left(4\right)}\leftarrow}=\delta_{u}^{\leftarrow}\left(\rho_{A\left(BCD\right)}\right)-\delta_{u}^{\leftarrow}\left(\rho_{AB}\right)-\delta_{u}^{\leftarrow}\left(\rho_{AC}\right)-\delta_{u}^{\leftarrow}\left(\rho_{AD}\right),
\\
\triangle_{\delta_{u\left(A\right)}}^{^{\left(4\right)}\rightarrow}=\delta_{u}^{\rightarrow}\left(\rho_{A\left(BCD\right)}\right)-\delta_{u}^{\rightarrow}\left(\rho_{AB}\right)-\delta_{u}^{\rightarrow}\left(\rho_{AC}\right)-\delta_{u}^{\rightarrow}\left(\rho_{AD}\right).
\end{eqnarray}
We can provide an upper bound of these two polygamy deficits:
$\triangle_{\delta_{u\left(A\right)}}^{^{\left(4\right)}\leftarrow}\leq S\left(\rho_{A}\right)-\frac{1}{2}I\left(\rho_{AB}\right)-\frac{1}{2}I\left(\rho_{AC}\right)-\frac{1}{2}I\left(\rho_{AD}\right)
=\frac{1}{2}I\left(\rho_{ABC}\right)$.
Since $I\left(\rho_{ABC}\right)=I\left(\rho_{A(BC)}\right)-I\left(\rho_{AB}\right)-I\left(\rho_{AC}\right)=\triangle_{I_{\left(A\right)}}^{\left(3\right)}$,
we have
\begin{eqnarray}
\triangle_{\delta_{u\left(A\right)}}^{^{\left(4\right)}\leftarrow}\leq\frac{1}{2}I\left(\rho_{ABC}\right)=\triangle_{I_{\left(A\right)}}^{\left(3\right)},
\
\triangle_{\delta_{u\left(A\right)}}^{^{\left(4\right)}\rightarrow}\leq\frac{1}{2}I\left(\rho_{ABC}\right)=\triangle_{I_{\left(A\right)}}^{\left(3\right)}.
\end{eqnarray}
When mutual information is polygamy $I\left(\rho_{ABC}\right)\leq0$,
we must have $\triangle_{\delta_{u\left(A\right)}}^{^{\left(4\right)}\leftarrow}\leq0$.

So far, similar as the polygamy deficit for tripartite states, we have defined the
two kinds of polygamy deficits for 4-partite pure states. We have provided
an upper bound for these two polygamy deficit, which is equivalent
to the interaction information for its tripartite reduced state $\rho_{ABC}$.
In general, the $I\left(\rho_{ABC}\right)$ can be positive or negative.
When $I\left(\rho_{ABC}\right)\leq0$, we have that the two kinds of polygamy
inequalities hold for one way discord. It is worth noting that  $\frac{1}{2}I\left(\rho_{ABC}\right)=\triangle_{I_{\left(A\right)}}^{\left(3\right)}$
holds for any tripartite reduced state $\rho_{ABC}$. That is to say,
when the mutual information is polygamy for $\rho_{ABC}$, we must
have $I\left(\rho_{ABC}\right)\leq0$, then both polygamy inequalities
of one way discord hold for $\rho_{ABCD}$. So we can make the polygamy
inequalities hold by adjusting the tripartite interaction information
$I\left(\rho_{ABC}\right)$ or the corresponding mutual information.
Since $I\left(\rho_{ABC}\right)$ does not include any measurement,
the experiment and calculations are very simple.

\begin{figure}[t]
\centering
\subfigure[]{\includegraphics[width=0.225\textwidth]{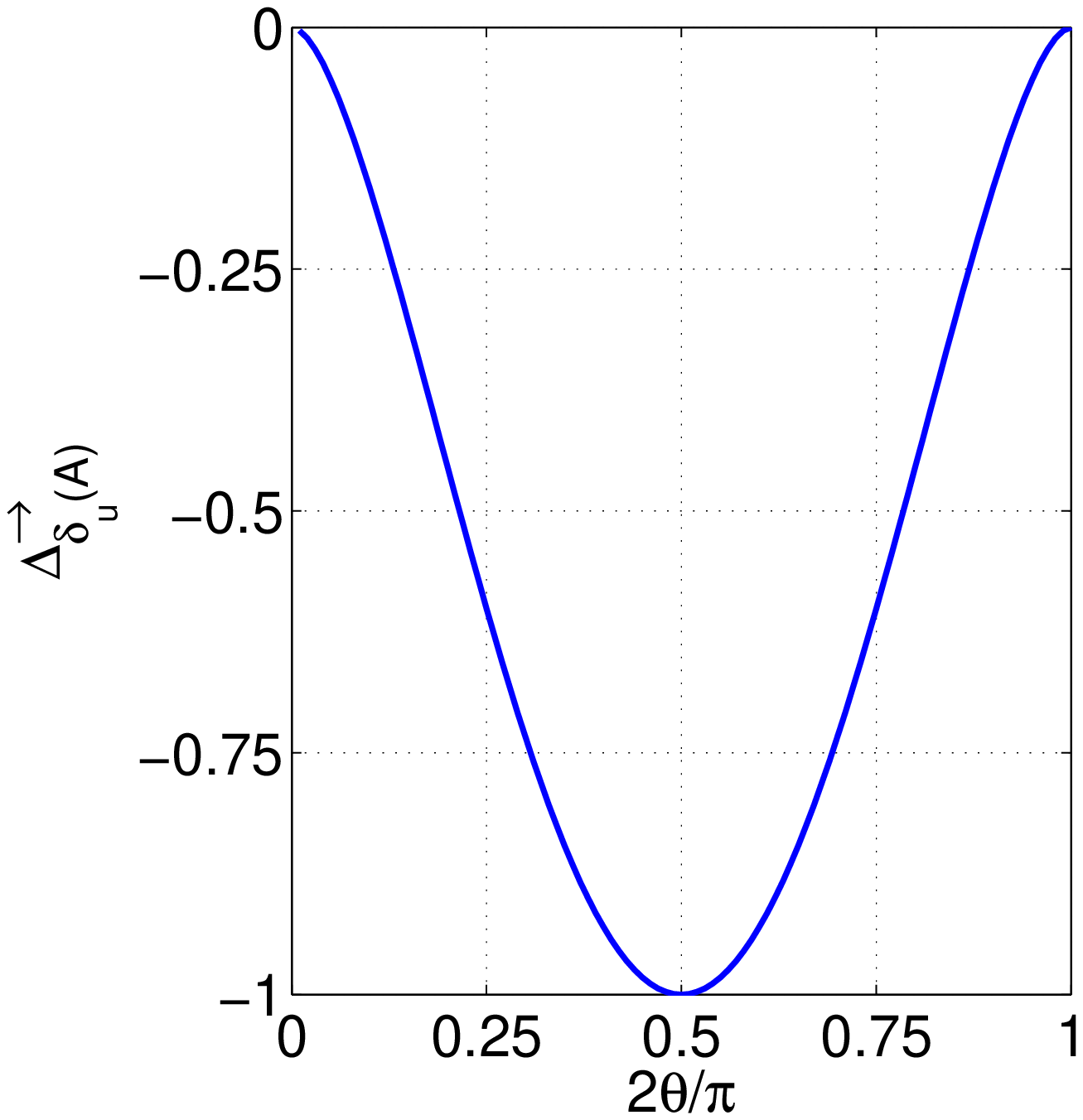}}
\subfigure[]{\includegraphics[width=0.22\textwidth]{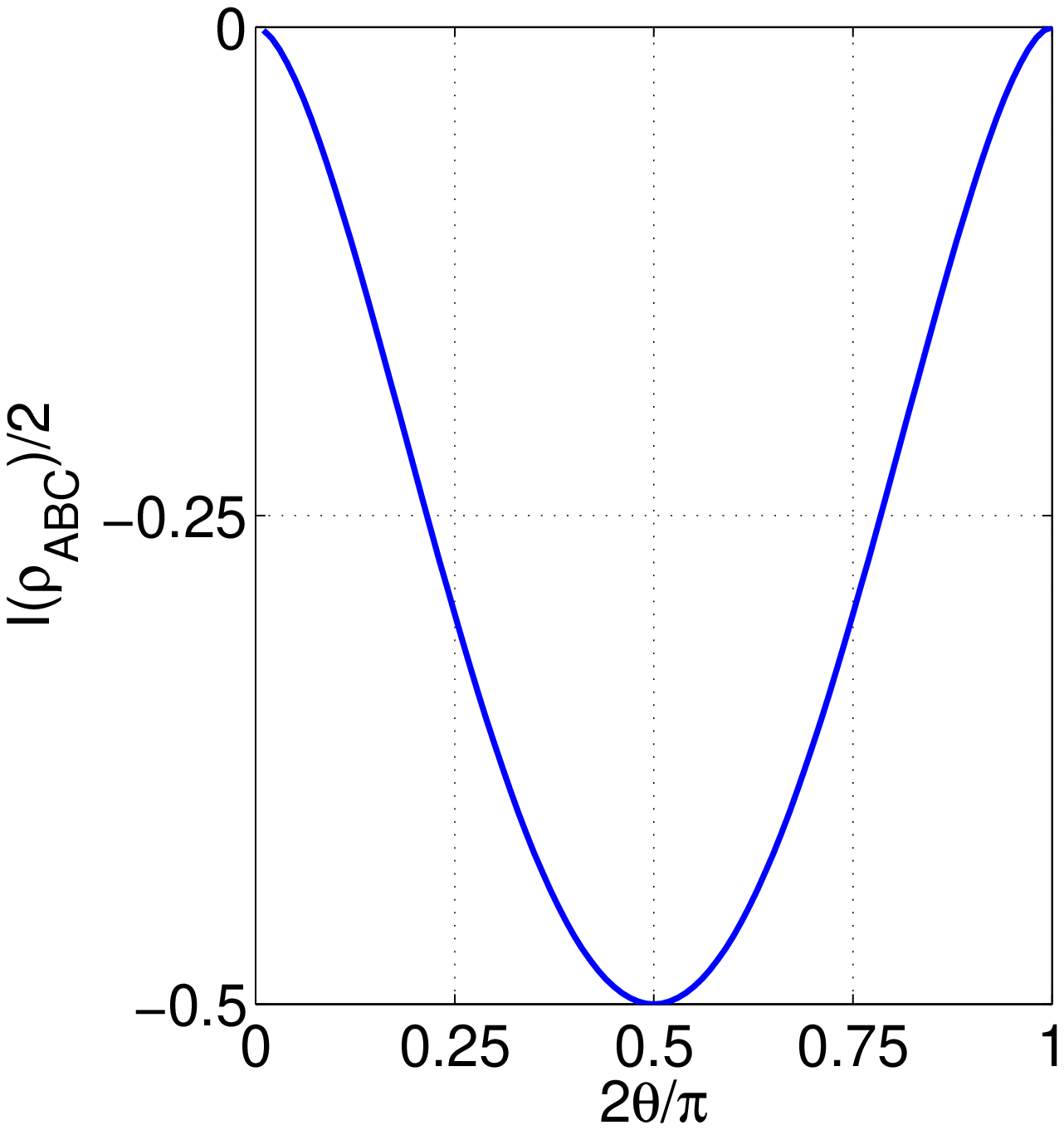}}
\caption{(color online). (a) For state $|\psi\rangle=\cos\theta|000\rangle+\sin\theta|111\rangle$
($\theta\in\left[0,{\pi}/{2}\right]$), the $\triangle_{\delta_{u\left(A\right)}}^{\rightarrow}$ against ${2\theta}/{\pi}$.
(b) The interaction information $I(\rho_{ABC})$ is plotted as a function of ${2\theta}/{\pi}$ for $|\psi\rangle=\cos\theta|0000\rangle+\sin\theta|1111\rangle$}
\label{f2}
\end{figure}

For example, we consider a family of states, $|\psi\rangle=\cos\theta|0000\rangle+\sin\theta|1111\rangle$
($\theta\in\left[0,{\pi}/{2}\right]$). In Fig. 2(b), the half of
the interaction information is plotted as a function of ${2\theta}/{\pi}$.
This figure shows that $\frac{1}{2}I\left(\rho_{ABC}\right)$
first decreases then increases with increasing $\theta$. In particular,
when $\theta=0$ or ${\pi}/{2}$, the state is separable and
the polygamy deficits are upper bounded by 0. When $\theta$= ${\pi}/{4}$,
the state is the maximally entangled state, the polygamy deficits are
upper bounded by -${1}/{2}$. That is to say, the one way discord
is always polygamy for this state.

\section{conclusions and discussion}
In summary, we have considered the distribution property of one way
discord for multipartite quantum systems.
We have showed that the difference between one way discord and quantum discord equals to
the difference between entanglement of assistance and entanglement of formation.
The distribution property of one way discord for tripartite pure state have also been
investigated. Moreover, we have introduced the concepts
of two polygamy deficits for one way discord of which we have found the equivalent
expressions. Using this result, we can obtain physical interpretations for these two polygamy deficits
and control the polygamy degree of one way discord regardless of the optimal measurement problem.
For 4-partite pure states, we have provided an condition for the case that one way discord is polygamy.
That is to say, if our condition is satisfied, the one way discord must be polygamy.
We believe that our results provide a powerful but computationally simplified method
in understanding the distribution property of one way discord in multipartite
quantum systems. Our results may have applications in quantum information
processing and can be applied to physical models of many-body quantum
systems.

\begin{acknowledgments}
We thank Yu Zeng and Xian-Xin Wu for valuable discussions. This work
is supported by ``973''
program (2010CB922904), NSFC (11075126, 11031005, 11175248) and NWU
graduate student innovation funded YZZ12083.
\end{acknowledgments}

\end{document}